\documentclass[cameraready]{Interspeech}

\title{Joint Learning Global-Local Speaker Classification to Enhance End-to-End Speaker Diarization and Recognition}

\author[affiliation={1,2}, equalcontribution]{Yuhang}{Dai}
\author[affiliation={2}]{Haopeng}{Lin}
\author[affiliation={2}]{Jiale}{Qian}
\author[affiliation={3}]{Ruiqi}{Yan}
\author[affiliation={2}]{Hao}{Meng}
\author[affiliation={1,2}]{Hanke}{Xie}
\author[affiliation={2}]{Hanlin}{Wen}
\author[affiliation={2}]{Shunshun}{Yin}
\author[affiliation={2}]{Ming}{Tao}
\author[affiliation={3}]{Xie}{Chen}
\author[affiliation={1}, correspondingauthor]{Lei}{Xie}
\author[affiliation={2}, correspondingauthor]{Xinsheng}{Wang}

\address{
    $^1$ Audio, Speech and Language Processing Group (ASLP@NPU), School of Computer Science \\ Northwestern Polytechnical University \\
    $^2$\text{Soul AI Lab, China} \\
    $^3$\text{X-LANCE Lab, Shanghai Jiao Tong University, China}
}

\email{yhdai@mail.nwpu.edu.cn, lxie@nwpu.edu.cn, wangxinsheng@soulapp.cn}


\usepackage{comment}
\usepackage{algorithm}
\usepackage{algorithmic}
\usepackage{amsmath}
\usepackage{multirow}
\usepackage{caption}
\usepackage{svg}

\begin{document}

\maketitle

\begin{abstract}
\vspace{-0.1cm}

Large Audio-Language Models (LALMs) have demonstrated remarkable performance in end-to-end speaker diarization and recognition. 
However, their speaker discriminability remains limited due to the scarcity of large-scale conversational data and the absence of explicit speaker representation optimization.
To address this, we propose GLSC-SDR, a paradigm that jointly trains speaker classification with diarization and recognition. We further introduce a Global-Local Speaker Classification strategy, which uses clustered speakers as global labels and re-encoded intra-cluster speakers as local labels. This hierarchical design enhances fine-grained speaker discrimination while preserving semantic transcription accuracy. Experiments on AliMeeting, AISHELL-4, and AMI-SDM demonstrate that GLSC-SDR achieves competitive or superior performance compared to simulation-based and multi-encoder approaches, without relying on large-scale real conversational data.
\end{abstract}

\vspace{-0.2cm}

\section{Introduction}
\vspace{-0.1cm}
Multi-speaker conversational transcription requires solving the fundamental problem of “who spoke what and when”, commonly referred to as Speaker Diarization and Recognition (SDR). Real-world conversations are characterized by rapid turn-taking, speaker overlaps, and acoustic variability, making robust speaker attribution a critical challenge~\cite{sd, park2022sdreview}.

Traditional SDR systems typically adopt a cascaded pipeline composed of Voice Activity Detection (VAD), Speaker Verification (SV), and Automatic Speech Recognition (ASR). Although modular and interpretable, such architectures suffer from error propagation and high system complexity. 
Recent advances in Large Audio-Language Models (LALMs) have demonstrated strong capabilities in unified end-to-end multi-speaker processing~\cite{hu2026tellwhisper, trainshort}, enabling structured sequence generation that jointly models timestamps, speaker labels, and transcripts~\cite{lalm_sd01, lalm_sd02}.

To enhance speaker attribution, several approaches extend the basic LALM framework with auxiliary components. For example, TagSpeech~\cite{tagspeech} incorporates an additional speaker encoder together with a time-anchoring strategy to improve fine-grained timestamp and speaker label prediction. SpeakerLM~\cite{yin2025speakerlm} introduces a speaker registration mechanism to strengthen identity consistency across segments.
Other works explore post-processing strategies~\cite{wang2024diarizationlm, post_process4sd}, such as leveraging large language models to refine diarization outputs and improve transcript readability~\cite{wang2024diarizationlm}.
While these efforts demonstrate that LALMs can be adapted to improve multi-speaker transcription, most approaches rely on architectural augmentation or inference-time refinement rather than explicitly optimizing speaker representation learning within the model. Consequently, speaker identity modeling remains largely implicit, and the learned representations may lack sufficient intra-speaker compactness and inter-speaker separability—particularly when conversational training data are limited or when acoustically similar speakers must be distinguished.

\begin{figure*}[t]
  \centering
  \vspace{-0.5cm}
  \includegraphics[clip, trim=0cm 6.1cm 0cm 6.3cm,width=\textwidth]{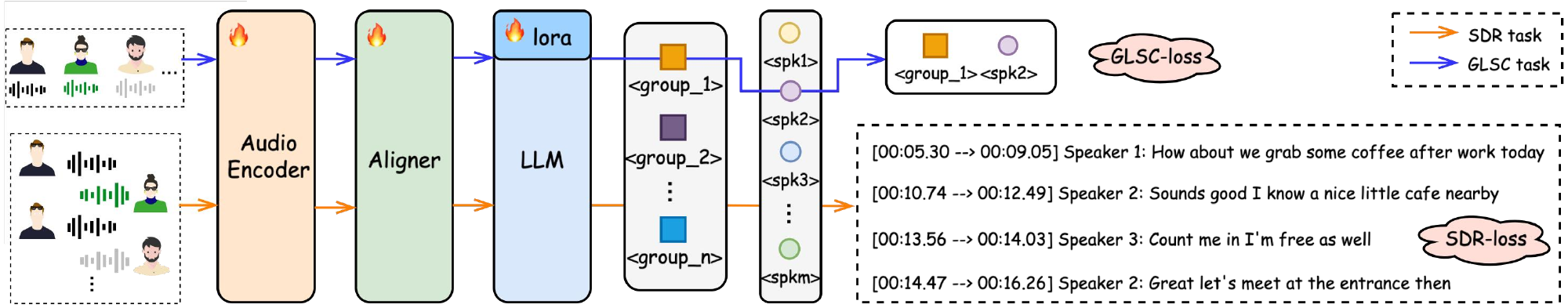} 
  \caption{Joint training Global-Local Speaker Classification and Speech Diarization and Recognition paradigm.}
  \captionsetup[figure]{skip=5pt} 
  \label{fig:glsc-sdr}
  \vspace{-0.6cm}
\end{figure*}

To overcome the limited speaker discriminability of current end-to-end models, we draw inspiration from the evolution of the Speaker Verification (SV) systems. 
Over the past decade, SV has evolved from statistical modeling~\cite{gmm_ubm, i_vector} to deep representation learning~\cite{x_vectors}. 
Building upon this paradigm, recent models further introduce metric learning and margin-based objectives within architectures such as ResNet and ECAPA-TDNN~\cite{ecapa_tdnn} to explicitly structure the embedding space. As a result, modern SV systems learn speaker representations with improved intra-class compactness and inter-class separability, enabling strong generalization to unseen speakers.


Inspired by these advances, we argue that although LALMs achieve strong performance in speech recognition and semantic modeling due to their powerful backbones~\cite{lalm_asr01, lalm_asr02}, their speaker representation learning remains largely implicit. As a result, the models may struggle to distinguish acoustically similar speakers in complex conversational scenarios.
To address this limitation, we introduce explicit speaker-aware supervision into LALMs by integrating speaker classification objectives into the training process. Specifically, we propose a new paradigm termed \textbf{G}lobal-\textbf{L}ocal \textbf{S}peaker \textbf{C}lassification jointly trained with \textbf{S}peaker \textbf{D}iarization and \textbf{R}ecognition (GLSC-SDR). By combining hierarchical speaker classification with the SDR objective, the proposed approach enhances the speaker discriminability of LALMs while preserving their end-to-end modeling capability, enabling more accurate multi-speaker transcription and tracking in conversational speech.
The main contributions of this paper are summarized as follows:
\begin{itemize}
\item We propose \textbf{GLSC-SDR}, a fully end-to-end joint training paradigm that tightly integrates speaker classification with the SDR task. Given multi-speaker audio as input, it directly produces structured outputs containing timestamps, speaker labels, and corresponding speech transcripts within a unified architecture.
\item We introduce a novel \textbf{GLSC} strategy tailored for LALMs. By incorporating hierarchical clustering and a speaker re-encoding mechanism, it provides both global and local speaker labels, which significantly enhance cross-segment speaker discriminability at the input representation level while requiring no architectural modification to the LLM backbone.
\item Extensive experiments on multi-speaker benchmarks, including AliMeeting, AISHELL-4, and AMI-SDM, demonstrate that the proposed paradigm substantially improves speaker diarization and recognition performance in complex conversational scenarios.
\end{itemize}

\vspace{-0.2cm}
\section{Method}
\vspace{-0.1cm}

Effectively enhancing a model’s ability to discriminate among different speakers is critical for the SDR task.
The core innovation of our approach lies in the integration of a hierarchical Global-Local Speaker Classification mechanism, which is jointly optimized with the Speaker Diarization and Recognition task. Diverging from traditional flat classification approaches, GLSC adopts a ``macro-acoustic clustering followed by micro-individual identification" strategy. This compels the model to simultaneously acquire cross-segment global consistency and fine-grained local discriminability at the input representation level. Such a design not only effectively mitigates the risk of overfitting in large-scale classification through an implicit curriculum learning effect but also significantly enhances the model's sensitivity to acoustic differences in complex multi-speaker scenarios. Notably, this paradigm achieves substantial performance gains solely through joint multi-task constraints without requiring additional network modules, as illustrated in Figure~\ref{fig:glsc-sdr}.
\vspace{-0.2cm}
\subsection{Problem Formulation}
\vspace{-0.1cm}
The goal of our GLSC-SDR framework is to improve the performance of SDR on multi-speaker audio. Given a raw waveform $S$, SDR aims to generate a structured sequence $O=(o_1, o_2, ..., o_l)$ containing timestamps, speaker labels, and transcriptions. While standard LALMs can model this sequence, their inherent speaker discriminability is often limited, particularly in complex multi-speaker scenarios.

To enhance SDR, we introduce the GLSC mechanism as a specialized training task within the LLM. GLSC generates a sequence of speaker labels for a single audio segment, where each label consists of a global speaker cluster label $L_g$ and a local speaker identifier $L_u$ within that cluster. 
This task explicitly teaches the model to recognize fine-grained speaker differences and cluster-level consistency, which indirectly improves the speaker attribution capability of the SDR task.
During training, the GLSC and SDR tasks are both serialized into unified sequence formats, with a special task token distinguishing them. Notably, GLSC sequences are single-utterance samples rather than full conversational data. The combined training objective is defined as:
\begin{equation}
L = \alpha L_{GLSC} + (1-\alpha) L_{SDR}
\end{equation}
where $L_{GLSC}$ and $L_{SDR}$ denote the losses for the GLSC and SDR tasks, respectively, and $\alpha$ serves as the weighting coefficient. During inference, the model adheres to the Serialized Output Training (SOT)~\cite{sot} paradigm, generating a structured sequence containing timestamps, speaker labels, and transcriptions.

\begin{table*}[ht]
\caption{Main results in Alimeeting, AISHELL-4 and AMI-SDM. (* indicates the results are coverted from~\cite{peng2026vibevoice})}
\vspace{-0.3cm}
\label{table:main_results}
\setlength{\tabcolsep}{4pt} 
\centering
\renewcommand{\arraystretch}{1.2} 
    \begin{tabular}{lcccccccccccc}
    \hline
    \multirow{2}{*}{Model}        & \multicolumn{4}{c}{Alimeeting}         & \multicolumn{4}{c}{AISHELL-4} & \multicolumn{4}{c}{AMI-SDM}  \\ \cline{2-13} 
    & WER↓ & cpWER↓ & $\triangle$cp↓ & SCA↑ & WER↓  & cpWER↓ & $\triangle$cp↓  & SCA↑ & WER↓  & cpWER↓  & $\triangle$cp↓ & SCA↑    \\ \hline
    \multicolumn{1}{l}{\textbf{Related work}}     &   &   &  &   &         &   &  & &         &         &         &         \\ \hline
    TagSpeech~\cite{tagspeech}  & 25.42          & 33.84          & 8.42          & \underline{81.63}          & -       & - & -& -            & 31.62   & 42.55   & 10.93   & 70.01   \\
    VibeVoice-ASR~\cite{peng2026vibevoice}  & 27.4           & 29.33          & \textbf{1.93} & - & \underline{21.4}           & \underline{24.99}    & 3.59          & -            & 24.65   & 28.82   & \textbf{4.17}        & -       \\
    Gemini-2.5-pro*  & 27.43          & 41.64          & 14.2          & - & 22.42   & 31.59          & 9.17          & -            & 22.35   & 34.78   & 12.43   & -       \\
    Gemini-3-pro*  & 26.75          & 32.84          & 6.09          & - & 22.75   & 27.43          & 4.68          & -            & 22.09         & \underline{26.91}          & \underline{4.82}           & -       \\ \hline
    \multicolumn{1}{l}{\textbf{Baseline}}   &   &   &  &   &         &   &  & &         &         &         &         \\ \hline
    Qwen2.5-omni~\cite{Qwen2.5-Omni}  & 29.21 & 43.64 & 14.43 & 71.28 & 31.46  & 46.28 & 14.82 & 76.68 & 32.25  &   49.86  & 17.61 & 60.06  \\
    Qwen2.5-omni-sft  & \underline{20.22}    & \underline{26.77}          & 6.55          & 81.28          & 23.83   & 26.34          & \underline{2.51}          & \underline{89.93} & \underline{20.23}   & 27.16   & 6.93    & \underline{74.33}   \\ \hline
    \multicolumn{1}{l}{\textbf{Our proposed}}  &   &   &  &   &         &   &  & &         &         &         &         \\ \hline
    GLSC-SDR & \textbf{20.09} & \textbf{25.43} & \underline{5.34} & \textbf{83.26} & \underline{\textbf{21.36}} & \textbf{23.49} & \textbf{2.13}    & \textbf{90.96} & \textbf{17.49} & \textbf{23.32} & 5.83    & \textbf{76.67}    \\ \hline
    \end{tabular}
\vspace{-0.5cm}
\end{table*}

\vspace{-0.2cm}
\subsection{GLSC}
\vspace{-0.1cm}


Conventional Speaker Diarization approaches primarily rely on local pairwise discrimination, often lacking a cross-segment global perspective. This limitation results in representations that are insufficiently robust in open scenarios. Conversely, naive global speaker classification typically treats each speaker ID as an independent category (i.e., flat classification). When confronted with highly similar voices or noisy labels, such methods are prone to overfitting and yield representation spaces with poor generalization capabilities. Consequently, constructing a globally consistent speaker representation space while simultaneously mitigating overfitting and effectively distinguishing between similar individuals has become a critical bottleneck in end-to-end SDR systems.

To address these challenges, we propose the Global-Local Speaker Classification (GLSC) framework. This framework is designed to overcome the limitations of the restricted local perspective in traditional SDR and to resolve the overfitting issues inherent in simple global classification when dealing with massive or acoustically similar speakers.

Diverging from traditional flat strategies that treat speaker IDs as isolated categories, GLSC adopts a hierarchical modeling paradigm characterized by ``macro-clustering followed by micro-identification." Specifically, we decompose the learning objective into a two-level structure. First, Global Coarse-grained Labels, generated via unsupervised clustering based on acoustic features, are employed to guide the model in capturing intrinsic acoustic commonalities (e.g., timbre and prosody), thereby constructing a robust and noise-tolerant feature space. Subsequently, within each global ``Super-Class," Local Fine-grained Labels are utilized to distinguish individual speakers, ensuring that the model retains the sensitivity to discriminate subtle differences among similar individuals after mastering the macro-classification. This hierarchical design organically integrates the strengths of unsupervised clustering and supervised identification, establishing an implicit Curriculum Learning mechanism. This not only prevents the model from overfitting to individual details during the early stages of training but also significantly enhances the generalization capability and discriminability of the representation space in open scenarios.


\vspace{-0.2cm}
\subsection{GLSC Data Construction Pipeline}
\vspace{-0.1cm}
\begin{figure}[t]
  \centering
  \vspace{-0.2cm}
  \includegraphics[clip, trim=6cm 0.4cm 3.5cm 0cm,width=0.5\textwidth]{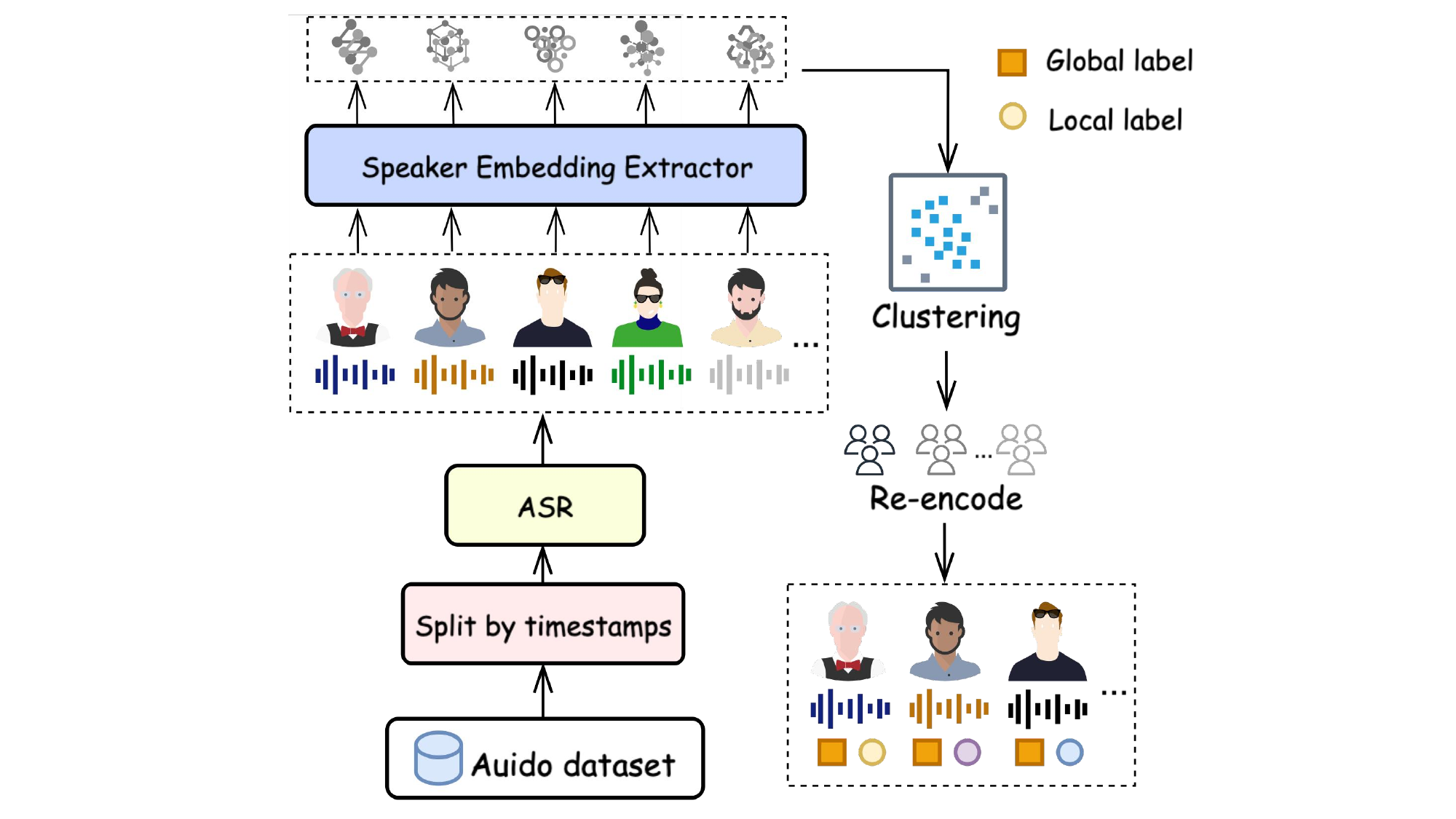} 
  \vspace{-0.6cm}
  \caption{GLSC data construction pipeline.}
  \captionsetup[figure]{skip=1pt} 
  \label{fig:glsc-pipeline}
  \vspace{-0.8cm}
\end{figure}

To support the training of the GLSC mechanism, we design a rigorous data preprocessing and label generation pipeline as illustrated in Figure~\ref{fig:glsc-pipeline}. First, we segment the audio based on the original dataset's timestamps to extract non-overlapping single-speaker speech segments. Next, an ASR model is introduced for quality filtering. Specifically, segments with a Word Error Rate (WER) exceeding 30\% or more than 2 insertion errors are identified as potential overlapping speech or crosstalk and are subsequently discarded as noisy data. Following this, a pre-trained speaker model is utilized to extract speaker embeddings from the remaining high-quality segments.

During the hierarchical label construction phase, we apply the HDBSCAN~\cite{hdbscan} algorithm for the unsupervised clustering of all speaker embeddings, merging clusters whose centroid cosine similarity exceeds 0.75. After filtering out outlier noise, the resulting macro-cluster IDs serve as global labels $L_g$ for the Global Speaker Classification (GSC) task. Concurrently, to facilitate the fine-grained modeling of acoustically similar speakers, we apply sequential encoding to distinct speakers within each cluster to generate local labels $L_u$. Finally, $L_g$ and $L_u$ are concatenated to construct composite GLSC labels for joint multi-task optimization.

\vspace{-0.2cm}
\section{Experiments}
 \vspace{-0.1cm}
\subsection{Experimental Setup and Implementation Details}
\vspace{-0.1cm}
\subsubsection{Datasets}
\vspace{-0.1cm}
We evaluate the proposed method on three widely used multi-speaker meeting benchmarks: AMI~\cite{ami}, AliMeeting~\cite{alimeeting}, and AISHELL-4~\cite{aishell4}. AMI is an English meeting corpus, while AliMeeting and AISHELL-4 are Mandarin far-field meeting datasets.

For AMI, we adopt the Single Distant Microphone (SDM) subset. For AliMeeting and AISHELL-4, we use the first channel from the 8-channel far-field microphone array recordings. All models are trained and evaluated independently within each dataset to ensure fair comparisons across different benchmarks.

For data preprocessing, following previous studies~\cite{kanda2022transcribe, tagspeech}, we adopt a turn-group-based audio segmentation strategy. Specifically, consecutive speaker turns without temporal gaps are merged into a single audio segment to construct training samples. This strategy helps preserve conversational coherence within each input sequence.

\vspace{-0.2cm}
\subsubsection{Model Configuration and Training Details}
\vspace{-0.1cm}
For speaker representation extraction, we employ ERes2Net~\cite{eres2net} as the speaker embedding model. As the backbone LALM, we adopt Qwen2.5-Omni-7B~\cite{Qwen2.5-Omni}, which provides strong audio-language alignment capability and supports long-context reasoning, making it suitable for complex multi-speaker conversational scenarios.

During training, we apply Low-Rank Adaptation (LoRA) fine-tuning to the AudioEncoder, Aligner, and Thinker modules within the LALM backbone. In the main evaluation experiments, the model is trained for 30 epochs to maintain consistency with the training configuration used in TagSpeech, enabling fair comparison. For ablation studies, the model is trained for 3 epochs. The initial learning rate is set to $1\times10^{-4}$, and the LoRA rank is set to 8.





\vspace{-0.2cm}
\subsection{Evaluation Metrics}
\vspace{-0.1cm}
The SDR task requires the system to generate accurate transcripts while correctly attributing each utterance to the corresponding speaker. To evaluate these capabilities, we adopt the following metrics.

\begin{itemize}
\item  \textbf{Word Error Rate (WER)}: WER is computed by concatenating all utterances in chronological order while ignoring speaker identities. This metric evaluates the pure speech recognition performance of the system.
\item  \textbf{Concatenated minimum-Permutation Word Error Rate (cpWER)}: cpWER evaluates the joint correctness of speech recognition and speaker attribution. It is calculated by first concatenating all utterances belonging to the same speaker and then finding the optimal permutation between the predicted and reference speakers that minimizes the WER~\cite{cpWER_watanabe2020chime}.
\item  \textbf{The difference between cpWER and WER ($\Delta cp$)}: The difference between cpWER and WER reflects the additional errors introduced by incorrect speaker attribution. Therefore, $\Delta cp$ serves as an indicator of speaker diarization performance independent of transcription errors.
\item  \textbf{Speaker Count Accuracy (SCA)}: SCA Measures whether the predicted number of speakers exactly matches the ground truth speaker count.
\end{itemize}

\section{Results and Analysis}

This section evaluates the proposed GLSC-SDR framework through extensive experiments. We first compare our approach against state-of-the-art (SOTA) systems to demonstrate its superior speaker attribution and transcription accuracy. Subsequently, we conduct ablation studies to validate the effectiveness of the hierarchical classification strategy and analyze the impact of clustering dynamics on model performance. Our results underscore the efficiency of GLSC-SDR in balancing semantic fidelity with robust speaker discrimination.

\subsection{Main results}

Table~\ref{table:main_results} summarizes the performance of GLSC-SDR against several competitive baselines across three benchmarks: AliMeeting, AISHELL-4, and AMI-SDM. To establish a rigorous baseline, we fine-tuned the vanilla Qwen2.5-Omni on the corresponding datasets (referred to as Qwen2.5-Omni-SFT) to evaluate the performance of standard supervised fine-tuning. Overall, GLSC-SDR achieves SOTA performance among end-to-end systems, demonstrating significant advantages in speaker attribution accuracy and system robustness.

Specifically, the vanilla Qwen2.5-Omni model exhibits poor performance on AliMeeting, confirming that unoptimized LALMs lack the necessary discriminative power for multi-speaker environments. While task-specific fine-tuning (Qwen2.5-Omni-SFT) improves results, its speaker attribution remains inferior to our proposed method. This indicates that simple fine-tuning is insufficient for robust speaker classification, whereas our GLSC strategy provides more targeted modeling.

Furthermore, GLSC strikes a favorable balance between semantic transcription (WER) and speaker attribution ($\triangle$cp, SCA). It preserves the semantic strengths of LALMs while reducing attribution errors by 18.47\% ($\triangle$cp relative to baseline) and surpassing TagSpeech in SCA—notably achieving this without relying on large-scale real-world dialogue data or complex multi-encoder architectures.





\vspace{-0.2cm}
\subsection{Analysis of classification methods}
\vspace{-0.1cm}

\begin{table}[]
\caption{The impact of different speaker classification methods on SDR effectiveness shown on Alimeeting.}
\vspace{-0.3cm}
\label{table:sc_method_ablation}
\renewcommand{\arraystretch}{1.2} 
\begin{tabular}{lcccc}
\hline
Classification Method       & WER↓           & cpWER↓         & $\triangle$cp↓          & SCA↑           \\ \hline
GSC-SDR  & \textbf{22.61}    & 30.81          & 8.2           & \underline{78.56}          \\
SC-SDR   & 22.83          & \underline{30.8}          & \underline{8.06}          & \underline{78.56}          \\ 
GLSC-SDR     & \underline{22.71}         & \textbf{29.93} & \textbf{7.22} & \textbf{79.74} \\ \hline
\end{tabular}
\vspace{-0.5cm}
\end{table}

To further understand GLSC's effectiveness, we investigate two key questions: (1) can the GLSC strategy perform effectively if only one classification level (global or local) is retained, and (2) does direct classification using original speaker labels achieve comparable performance? 

As shown in Table~\ref{table:sc_method_ablation}, the experiments demonstrate that GLSC outperforms both Global-Only Classification (GSC) and direct speaker-label classification (SC). Global-only classification suffers from high speaker attribution errors, as indicated by the elevated $\triangle$cp value, because global clusters alone cannot resolve intra-cluster speaker ambiguities. Direct speaker-label classification achieves the poorest semantic transcription quality and highest WER, as it lacks global acoustic constraints, impairing cross-segment speaker consistency. In contrast, GLSC achieves the lowest $\triangle$cp and highest speaker count accuracy, confirming that the hierarchical ``global clustering first, local discrimination second" paradigm effectively balances cross-segment consistency with fine-grained speaker discriminability.

\subsection{Impact of clustering number}

\begin{table}[]
\centering
\caption{The impact of cluster number on GLSC and cluster algorithm, results shown on Alimeeting.}
\vspace{-0.3cm}
\label{table:ablation_cluster_num}
\setlength{\tabcolsep}{1pt} 
\renewcommand{\arraystretch}{1.2} 
\begin{tabular}{lccccc}
\hline
exp\_group    & cluster num & WER$\downarrow$  & cpWER$\downarrow$ & $\triangle$cp$\downarrow$ & SCA$\uparrow$  \\ \hline
baseline      & -                  &            23.33  &            32.18  &            8.85  &              77.84 \\ \hline
GLSC\_Kmeans  & 50                 &            23.05  &            31.53  &            8.48  &              72.2  \\ 
GLSC\_Kmeans  & 200                & \textbf{22.58} &    \underline{30.32} & \underline{7.74} &   \underline{78.58} \\
GLSC\_Kmeans  & 430                &            22.7  &            31.28  &            8.23  &              78.02 \\
GLSC\_HDBSCAN & 430                &    \underline{22.71} & \textbf{29.93} &    \textbf{7.22}  &      \textbf{79.74} \\ \hline
\end{tabular}

\end{table}

We further investigate how the choice of clustering method and the granularity of speaker groups affect GLSC-SDR’s performance.
As shown in Table~\ref{table:ablation_cluster_num}, 
both the number of clusters and the clustering algorithm exert a significant impact on GLSC performance. In particular, HDBSCAN—a density-based clustering algorithm—consistently outperforms K-means~\cite{kmeans} cross all metrics. This is because HDBSCAN naturally adapts to the non-uniform distribution of speaker features and effectively mitigates the influence of outliers, whereas K-means is sensitive to initial centroid placement and prone to intra-cluster feature mixing.

Cluster granularity also plays a critical role. Increasing the number of clusters does not necessarily improve performance: the optimal results are obtained with 200 clusters. Too few clusters lead to overcrowded groups, reducing fine-grained discriminability and increasing speaker attribution errors, while an excessive number of clusters creates intra-cluster redundancy, which hampers the model’s ability to learn discriminative representations and degrades cpCER.

\section{Conclusion}

In this work, we proposed GLSC-SDR, a Global-Local Speaker Classification paradigm jointly trained with the Speech Diarization and Recognition task, to enhance the speaker discriminability and transcription accuracy of LALM-based SDR systems. Extensive experiments on AliMeeting, AISHELL-4, and AMI-SDM demonstrate that GLSC-SDR consistently achieves competitive or superior performance compared to SOTA approaches, with notable improvements in speaker attribution and speaker count prediction. By leveraging a hierarchical global-local classification mechanism and a joint training strategy, our approach effectively reduces speaker attribution errors while preserving the semantic transcription capabilities of the underlying LALM. Importantly, these gains are realized without relying on large-scale real-world conversational data or complex multi-encoder architectures, highlighting GLSC-SDR as a practical, scalable, and efficient solution for end-to-end multi-speaker SDR tasks.


\newpage




\bibliographystyle{IEEEtran}
\bibliography{mybib}

\end{document}